\documentclass[twocolumn]{aastex62}
\usepackage{hyperref}
\usepackage{cleveref}
\received{...}
\revised{...}
\accepted{\today}
\submitjournal{ApJ}

\shorttitle{X-ray Source in NGC 3413}
\shortauthors{Sonbas, Dhuga, \& G\"o\u{g}\"u\c{s}}

\begin{document}

\title{The Bright X-Ray Source in NGC 3413}

\email{edasonbas@gmail.com}

\author{E. Sonbas}
\affil{University of Adiyaman, Department of Physics, 02040 Adiyaman, Turkey}
\affil{Department of Physics, The George Washington University, Washington, DC 20052, USA}

\author{K. S. Dhuga}
\affil{Department of Physics, The George Washington University, Washington, DC 20052, USA}

\author{E. G\"o\u{g}\"u\c{s}}
\affil{Sabanc\i~University, Orhanl\i~- Tuzla, Istanbul 34956, Turkey}



\begin{abstract}
\noindent
The emission-line dwarf galaxy NGC 3413 is known to host a bright X-ray source near its optical center. The 0.3-10 keV luminosity of this source is estimated to be approximately 10$^{39}$ erg$s^{-1}$ potentially qualifying it as an ultra-luminous X-ray (ULX) source. A recent XMM-Newton observation suggests that the source is not point-like, and instead, is more likely a composite of point-like sources with extended and/or diffuse emission. The spectral and temporal features of the bright region are similar to those associated with the so-called broadened disk state of ULXs. Based on a multi-color blackbody spectral fit, we estimate the mass of the bright source to be in the range 3 - 20M$_{\odot}$. Potential optical counterparts are also explored with the aid of available SDSS and  PanStars data.

\end{abstract}



\section{Introduction} \label{sec:intro}
\noindent
Many point-like X-ray sources with luminosities  above a threshold luminosity of L$_x$ $>$ 10$^{39}$ erg$s^{-1}$, have been found in nearby galaxies \citep{astro-ph/0307077, 2004ApJS..154..519S, 2005ApJS..157...59L, 2015ApJ...805...12L}. Assuming isotropic emission, some of these sources have luminosities well in excess of 10$^{40}$ erg$s^{-1}$, surpassing emission beyond the Eddington limit for a stellar mass BH with a mass in the range 3 - 20 $M_{\odot}$. These ultra-luminous X-ray sources (ULXs) are non-nuclear i.e., located off-center from the nucleus of the host galaxy and therefore unlikely to be powered by accretion onto the central super massive black hole (SMBH). Several possibilities as to the nature of these intriguing objects continue to be discussed in the literature: a) an early model, though now seemingly less likely, poses the existence of BHs in the intermediate mass range of M$\sim$10$^{2}$ - 10$^{5}$M$_{\odot}$, accreting at sub-Eddington rates \citep{1999ApJ...519...89C, 2004ApJ...607..931M}; b) current models, based on recent data, tend to lean toward stellar-mass BHs (sMBHs) with a possible combination of effects such as relativistic beaming, and/or accretion at super-Eddington limits (\citep{2001ApJ...552L.109K, 2002ApJ...568L..97B, 2007Ap&SS.311..203R} and references therein). Indeed, very recent evidence, the detection of pulsations in a handful of sources \citep{2014Natur.514..202B, 2016ApJ...831L..14F, 2017Sci...355..817I}, strongly argues in favor of at least a fraction of these sources hosting neutron stars, implying an overall heterogeneous underlying population as opposed to a single class of objects.\\ 
\\
ULX candidates have been found in diverse environments, including star-forming regions of spiral galaxies, elliptical galaxies, as well as dwarf galaxies. Dwarf galaxies are of special interest because of the existence of luminous X-ray sources in these less massive systems may have had a more significant impact on their early evolution in comparison with their more massive counterparts. \citet{2015ApJ...805...12L} with a 1.7-ks Chandra snapshot of NGC 3413 at 8.6 Mpc detected the presence of a bright X-ray source approximately 3$^{\prime\prime}$ from the optical center. Although \citet{2015ApJ...805...12L} were able to determine a luminosity of L$_{x}$ $\sim$1 x 10$^{39}$ erg$s^{-1}$, thus confirming the possible ULX candidate, the limited exposure lacked sufficient statistics to further elucidate the nature of the detected source. Of course, luminosities of the order of few $\times$ 10$^{38}$ erg$s^{-1}$ are also possible in XRBs containing (ordinary) stellar BHs and NS. The X-ray luminosity contributions from these type of sources can be estimated from the relations given by \citet{2010ApJ...724..559L}: the expected collective X-ray binary (XRB) luminosity can be parametrized in terms of the star formation rate and the contributions from LMXBs and HMXBs. For NGC 3413, the X-ray luminosity expected from the candidate ULX is in excess of the XRB contributions. Thus the source in NGC 3413, an emission-line dwarf galaxy with a mass of $\sim$10$^{8}$ M$_{\odot}$, relatively low SFR (0.06 $\pm$ 0.02 M$_{\odot}$$yr^{-1}$), and a low hydrogen column density (N$_H$$\sim$ 2$\times$10$^{20}$cm$^{-2}$), is potentially of high value as a prime ULX candidate. \\
\\
Recent studies indicate that ULXs exhibit different spectral behavior in comparison with low-luminosity galactic BHs, featuring two-component spectra with soft excess and a turnover at energies near 5 keV. Indeed, \citet{2013MNRAS.435.1758S} extracted fractional variability and constructed variability-hardness diagrams to distinguish three main states i.e., the broadened disk, hard ultraluminous, and the soft ultraluminous. One of the conclusions they reached was that ULXs with an X-ray luminosity of $<$ 3 $\times$ 10$^{39}$ erg s$^{-1}$ are dominated by broadened disk spectra. 
\\
\\
In this work, we present the results of a spectral analysis  of a recent  $\sim$ 100ks  $XMM-Newton$ observation of the ULX candidate in NGC 3413. Our results suggest that of the three recently identified ULX transition states \citep{2009MNRAS.397.1836G, 2013MNRAS.435.1758S}, the source in NGC 3413 falls into the broadened disk category. The paper is organized as follows: Section 2 reports the details of the observations and the data reduction; Section 3 presents the results of the spectral and temporal analysis of $XMM-Newton$ data, as well as, our findings into the inquiry of potential optical counterparts probed via a color-magnitude diagram (based on SDSS and PanStars data) and mass estimates for the secondary. In section 4, we summarize our findings.         
\section{Observations and Data Reduction}
\noindent
We observed NGC 3413 with the European Space Agency’s (ESA) X-ray Multi-Mirror Mission (XMM-Newton) for $\sim$ 100ks in two pointings in 2016: November 15 and 19-20 respectively (PI: K. S. Dhuga; see Table 1 for details). In Figure 1a, we present the X-ray image of the dwarf galaxy along with its 90\% optical brightness contour and optical center. We clearly observe that the bright X-ray source is off-center and non-point like i.e., the source is diffuse and/or comprises several point-like sources within an angular separation of $\sim$8$^{\prime\prime}$ (see Figure 1a). Using that as a working hypothesis, we separated the main parts of the emission into two regions indicated by $8^{\prime\prime}$ yellow circles. The brighter source region (lower left with respect to the optical center) is designated src1, and the other one (upper right relative to the optical center) is designated src2.\\
\begin{figure}
\plotone{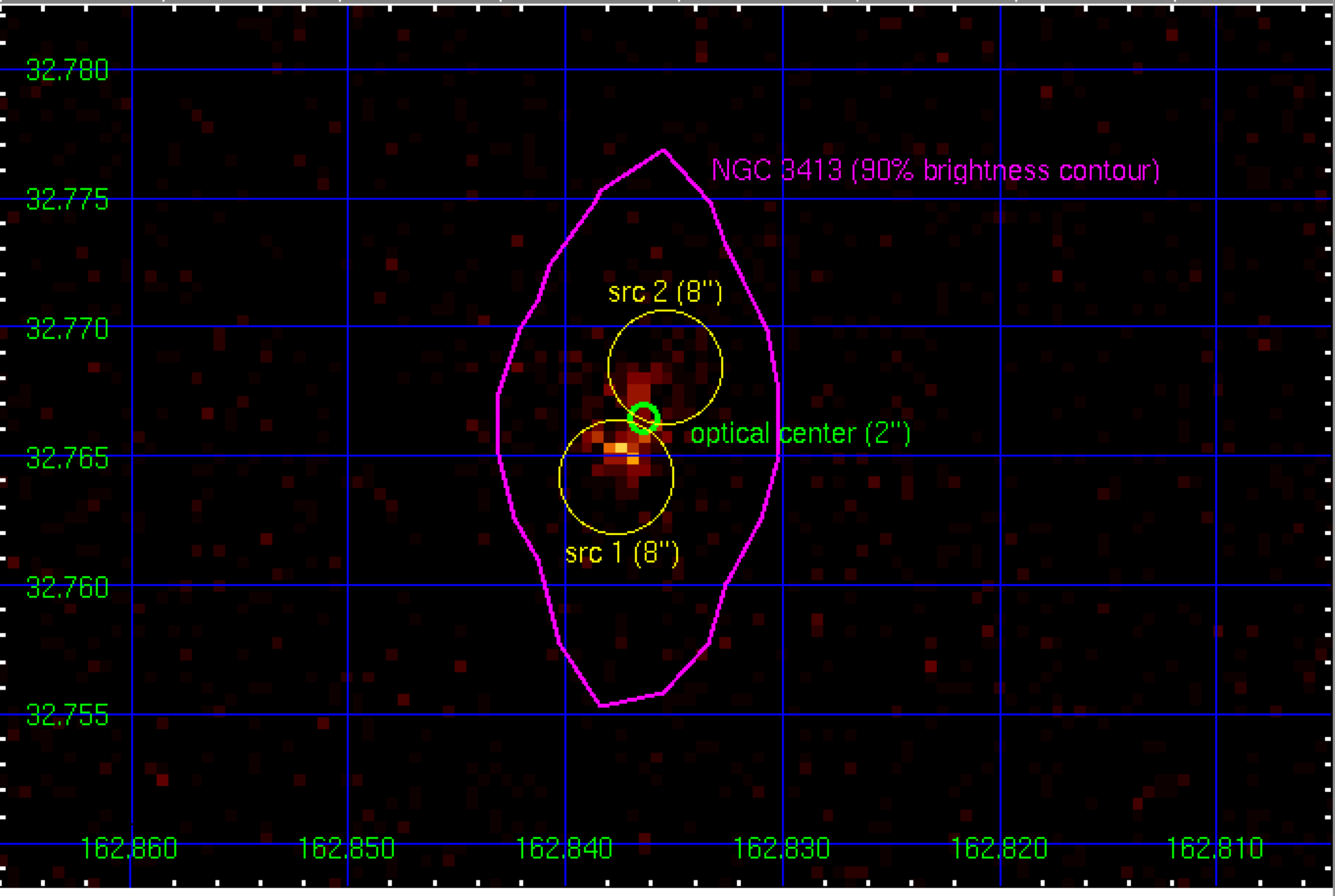}
\plotone{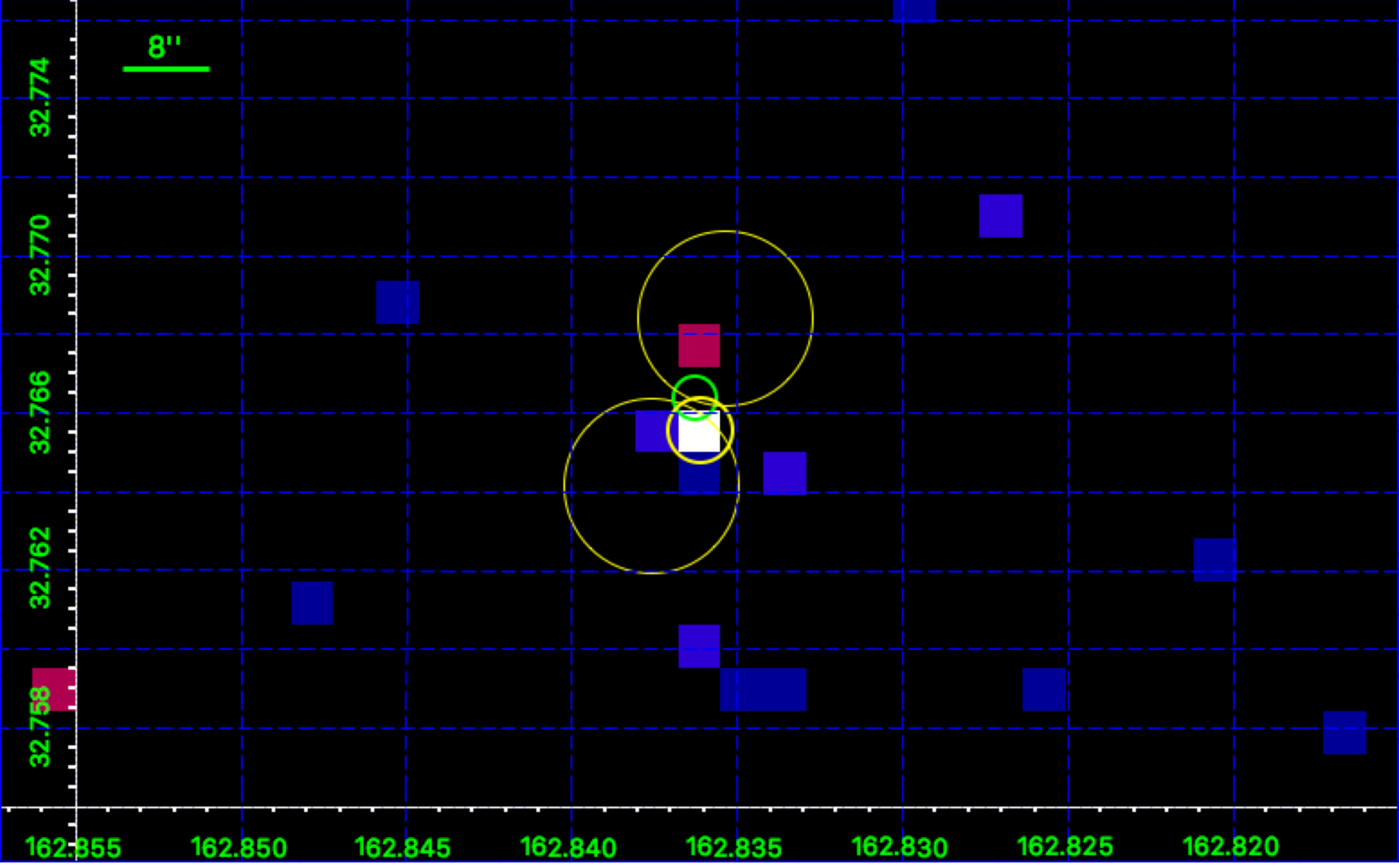}
\caption{{\bf a)} XMM-Newton image ($\sim$ 100 ks exposure) of the X-ray source in NGC 3413: The magenta outline is the 90$\%$-optical-brightness contour of NGC 3413. The yellow circles (8$^{\prime\prime}$) indicate src1 and src2 respectively. The green circle indicates the optical center. {\bf b)} Chandra image ($\sim$ 1.7 ks exposure) of the X-ray source (3$\arcsec$ yellow circle) in NGC 3413. \label{fig:fig1}}
\end{figure}
\begin{deluxetable}{crrc}
\tablecaption{X-ray observations of NGC 3413}
\tablehead{ \colhead{Date} & \colhead{ObsId} & \colhead{Exp} & \colhead{Effective Exp}\\
\colhead{} & \colhead{} & \colhead{(ks)} & \colhead{(ks)}}
\startdata
2016 November 15 & 0781410101 & 56 & mos2: 40 pn: 30 \\
2016 November 19 & 0781410201 & 53 & mos2: 50 pn: 35 \\
\enddata 
\label{table:nonlin} 
\end{deluxetable}
\\
\noindent For the X-ray source and for the diffuse background emission we extracted spectra and performed spectral analyses using the European Photon Imaging Camera (EPIC) data. The observation data files (ODFs) were processed using standard procedures of the \emph{XMM-Newton} Science Analysis System (XMM-SAS version 14.0.0). We excluded MOS1 data from our spectral analysis because the source for this detector was in the neighborhood of a detector chip gap. We extracted spectra using circular ($r=8''$) regions centered on the X-ray sources for each observation. Background subtraction was performed using source-free apertures located on the same chip. A standard event screening described in the \texttt{multixmmselect} manual\footnote{http://xmm.esac.esa.int/sas/current/documentation/threads/ multiexposures\_thread.shtml} was applied to appropriate GTI files to produce final data products. Finally, \texttt{epicspeccombine} was used to combine spectrum files of two observations to create one total spectrum file and the corresponding rmf, arf and bkg files.\\ 
\\
The spectral analysis of the X-ray data was performed using XSPEC version 12.9.1. The pn and MOS spectral energy channels were grouped to have at least 5 counts per bin. 
Spectra of the sources were fitted in the 0.3 - 10 keV energy range with a combination of power law, multicolor disk (diskbb), and emission of hot diffuse plasma (mekal and Raymond) models with N$_H$ allowed to vary. The $phabs$ model in XSPEC was used \citep{1992ApJ...400..699B} for the interstellar absorption. Best-fit parameters for each model can be found in Table 1, while the spectra and the fits are shown in Figures 2-3.\\

\section{Results and Discussion}
\noindent
We fitted the XMM spectra for src1 and src2 with a combination of spectral models appropriate for both point-like sources and those that tend to produce diffuse emission. The best fits to src1 and src2 are displayed in Figure 2; see the $upper$ and $lower$ panels respectively. The fit to src1 ($upper$ panel) is obtained with a combination of a relatively hot disk (1.9 $\pm$ 0.2 keV) and a steep power law ($\Gamma$ = 5.2 $\pm$ 0.9). Normally, these parameters would be considered unacceptable or at least highly suspicious for galactic stellar-mass BHs but on the contrary, for ULX spectra, a number of studies suggest the presence of a hot broad disk and a steep power law that acts to compensate for the soft photon excess. Thus in the case of ULX spectra these parameters are not necessarily unusual especially in the so-called broadened disk state \citep{2009MNRAS.397.1836G, 2013MNRAS.435.1758S}. For src2 an acceptable fit could only be obtained from a combination of diffuse emission from warm ($\sim$ 0.7 $\pm$ 0.1 keV) and relatively hot ($\sim$ 3.2 $\pm$ 0.4 keV) gas; the fit shown in the $lower$ panel of Figure 2 is obtained with a combination of mekal+Raymond models. Emission from a single temperature plasma model did not produce a meaningful fit.\\
\\
Although the (PL+ diskbb) fit to src1 is statistically acceptable and the residuals look reasonable, given the extended nature of the emission region near src1 in the XMM image, there is still the possibility of some contamination due to diffuse emission. We explored this possibility by adopting the following procedure: We analyzed the 1.7-ks Chandra observation of NGC 3413 by Lemons et al 2015. We show the Chandra image in Figure 1b; It strongly suggests the presence of a localized source, indicated by a 3$\arcsec$ yellow circle, presumably the point source in src1. The extracted spectrum from this region is dominated by counts above 1 keV i.e., a hard spectrum. This is consistent with the notion that the majority of the central emission is most likely from a point source in src1. To test whether the src1 emission contains a diffuse component, we took a two-step approach in our spectral analysis of the XMM data: Instead of using an 8” region for scr1, we selected a 2$\arcsec$ region centered on the Chandra position for src1 (and an appropriate background region of same size) and extracted a new spectrum. We found this spectrum to be best fitted with a single PL with an index of 1.6$\pm$0.3; the fit is shown in Figure 3 ($upper$ panel). We find this to be reassuring since the limited Chandra data indicate a hard spectrum for the central emission in this region. In the next step, we fitted the entire src1 region (i.e., 8$\arcsec$ as in original analysis) with a combination of PL (fixed parameters from the previous step) + Mekal model to account for any diffuse emission from a hot gas. This produces a statistically acceptable fit, very similar in quality (almost identical $\chi^2$) to our original fit with PL + diskbb and is shown in Figure 3 ($lower$ panel). The contribution from the Mekal is primarily in the soft band with a relatively low temperature (kT) of 0.14 $\pm$ 0.01 keV. While this potentially argues for some diffuse contribution (soft component) to src1, we note two observations which leads us to suspect this scenario: a) the extracted gas temperature is rather low especially compared with the more typical gas temperature of 0.7$\pm$0.1 keV needed to fit the src2 spectrum, and b) the fit requires us to fix the PL index i.e., setting the PL index as a free parameter produces a very poor fit. Thus overall, the fits (see Table 2 for the parameters) are consistent with our conjecture that the bright central source in NGC 3413 is most likely a combination of point-like and diffuse emission sources \citep{2011ApJS..192...10L}. The fit to src1 (the brighter portion of the central source) suggests the presence of a broadened (hot) disk state noted in other ULX studies. The steep power law apparently compensates for the soft photon excess. Both of these features are associated with ULX sources that have been studied with higher precision data \citep{2009MNRAS.397.1836G, 2013MNRAS.435.1758S}. We also tried to ascertain whether the spectrum for src1 displays any curvature. Modeling its spectrum with a broken power law, the resulting break energy, while not tightly constrained, is implied around $\sim$ 3 keV. The typical ULX spectra exhibit a break around $\sim$ 5 keV.\\
\begin{figure}
\plotone{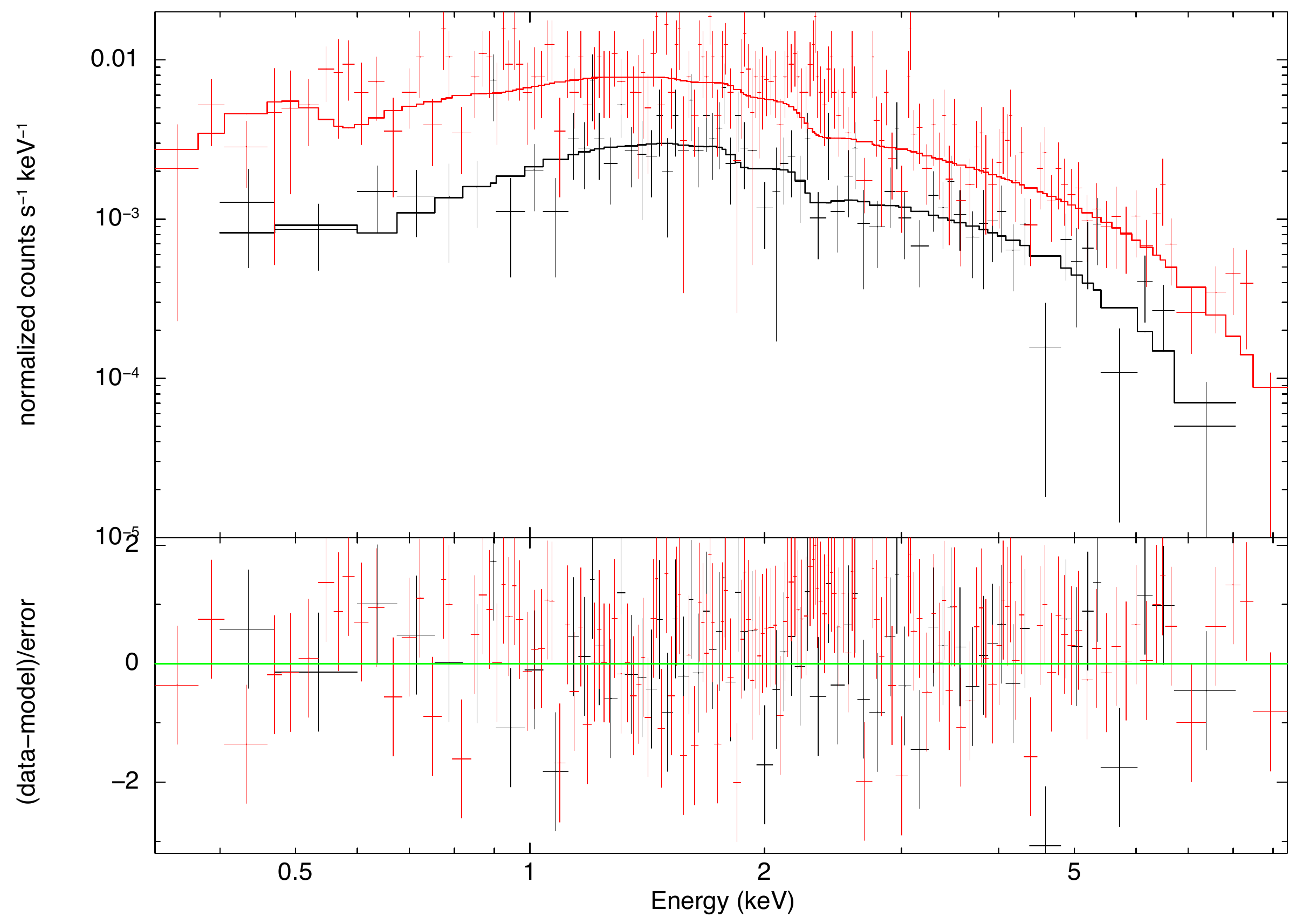}
\plotone{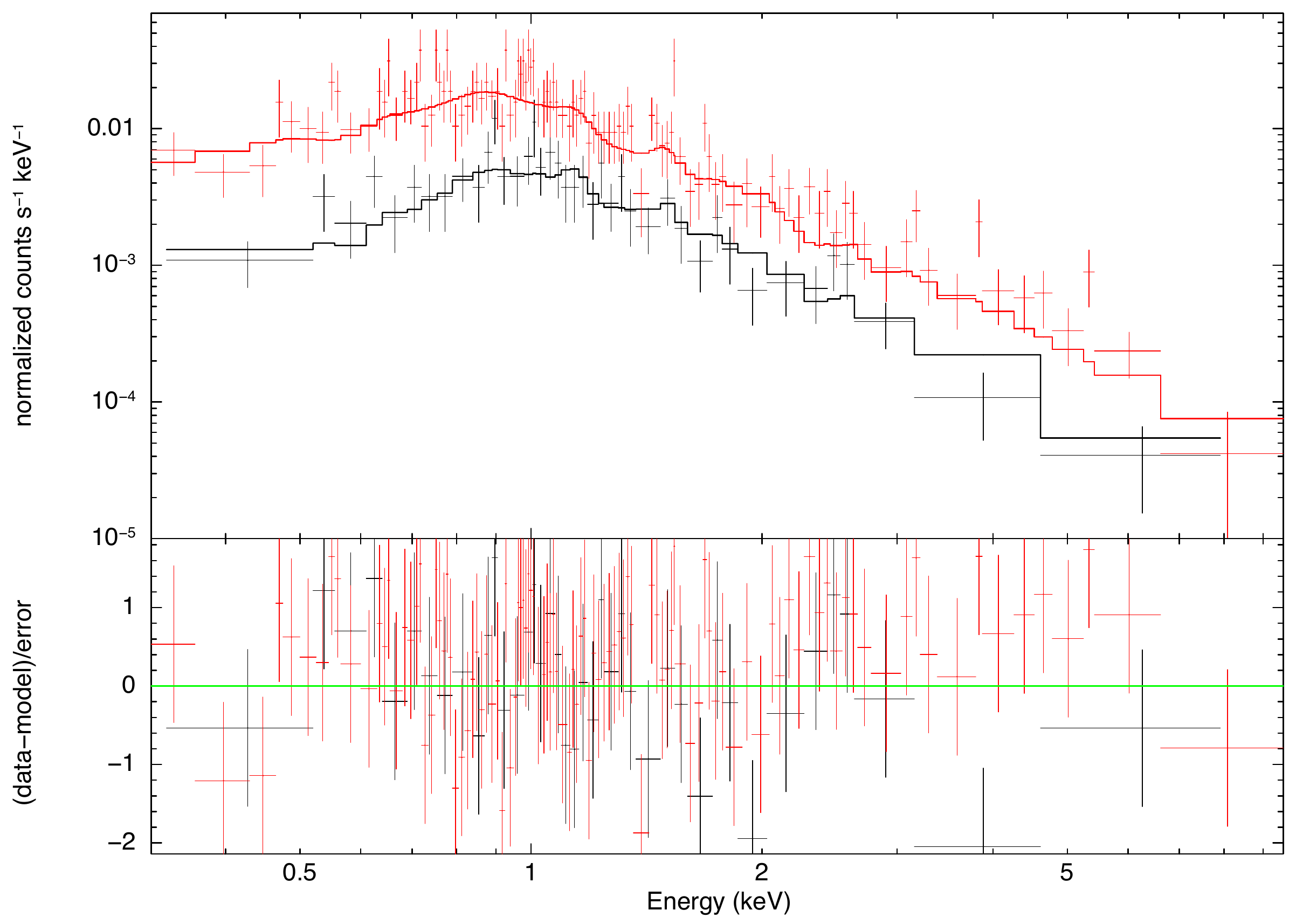}
\caption{($upper$) XMM-Newton EPIC pn and MOS2 spectra of src1 fitted with diskbb + PL, and ($lower$) XMM-Newton spectra of src2 fitted with mekal+Raymond models (see text for details). \label{fig:fig2}}
\end{figure}

\begin{deluxetable*}{crrrrrrrr}
\tablecaption{Summary of fit parameters of X-ray source(s) in NGC 3413}
\tablehead{ \colhead{Source Name} & \colhead{Model} & \colhead{N$_{H}$} &\colhead{Flux (0.5 - 10.0 keV)} & \colhead{$\Gamma$}  & \colhead{T$_{in}$}  & \colhead{kT} & \colhead{$\chi^{2}$/dof}  \\
\colhead{} & \colhead{} & \colhead{(10$^{22}$ cm$^{-2}$)} & \colhead{(erg cm$^{-2}$ s$^{-1}$)} & \colhead{} & \colhead{(keV)} &\colhead{(keV)}  & \colhead{} }
\startdata
src1 & PL+diskbb &0.39 & 1.0$^{+0.1}_{-0.3}$ $\times$ 10$^{-13} $ & 5.2 $\pm$ 0.9 & 1.9 $^{+0.4}_{-0.3}$ &\nodata& 202/220\\
src1 & mekal+PL &0.51 &  2.3$^{+0.2}_{-0.2}$ $\times$ 10$^{-13} $ &  1.64 (fixed)  & \nodata &0.14$\pm$0.01 & 213/217\\
src2 & mekal+Raymond & 0.08 & 4.0$\pm$0.5 $\times$ 10$^{-14}$ & \nodata & \nodata & 0.69 $^{+0.2}_{-0.1}$ / 3.2 $^{+0.6}_{-0.5}$ & 138/180 \\
\enddata 
\label{table:nonlin} 
\end{deluxetable*}
\noindent Using the diskbb normalization obtained for our spectral fit to src1, we can estimate the mass of the central object. Here we follow the procedure described by \citet{2000ApJ...535..632M}, who parametrized the inner radius (R$_{in}$) of the disk as R$_{in}$ = $\xi . \kappa^{2} . R_{eff}$, where $\xi_{i}$ (= 0.412)  and $\kappa$ (= 1.7) are correction factors taken from \citet{1995ApJ...445..780S} and \citet{1998PASJ...50..667K}, respectively. These factors primarily correct for the fact that the inner disk temperature does not necessarily peak at the inner radius R$_{in}$ and that the color temperature does not correspond to the effective temperature of the disk; R$_{eff}$ is the effective radius obtained directly from the diskbb normalization. We obtain a corrected radius (R$_{in}$)  of ~ (29 $\pm$5) km, which leads to a mass of $\sim$3M$_{\odot}$ for a non-rotating BH. If we assume a maximally spinning BH, the calculated mass is $\sim$20M$_{\odot}$. Clearly the outcome depends on a number of sensitive assumptions and as a result the mass range varies by an order of magnitude but nonetheless the exercise still suggests that the compact object, if a BH, is more likely to be in the stellar mass range as opposed to the more massive category of intermediate mass BHs.\\
\begin{figure}
\plotone{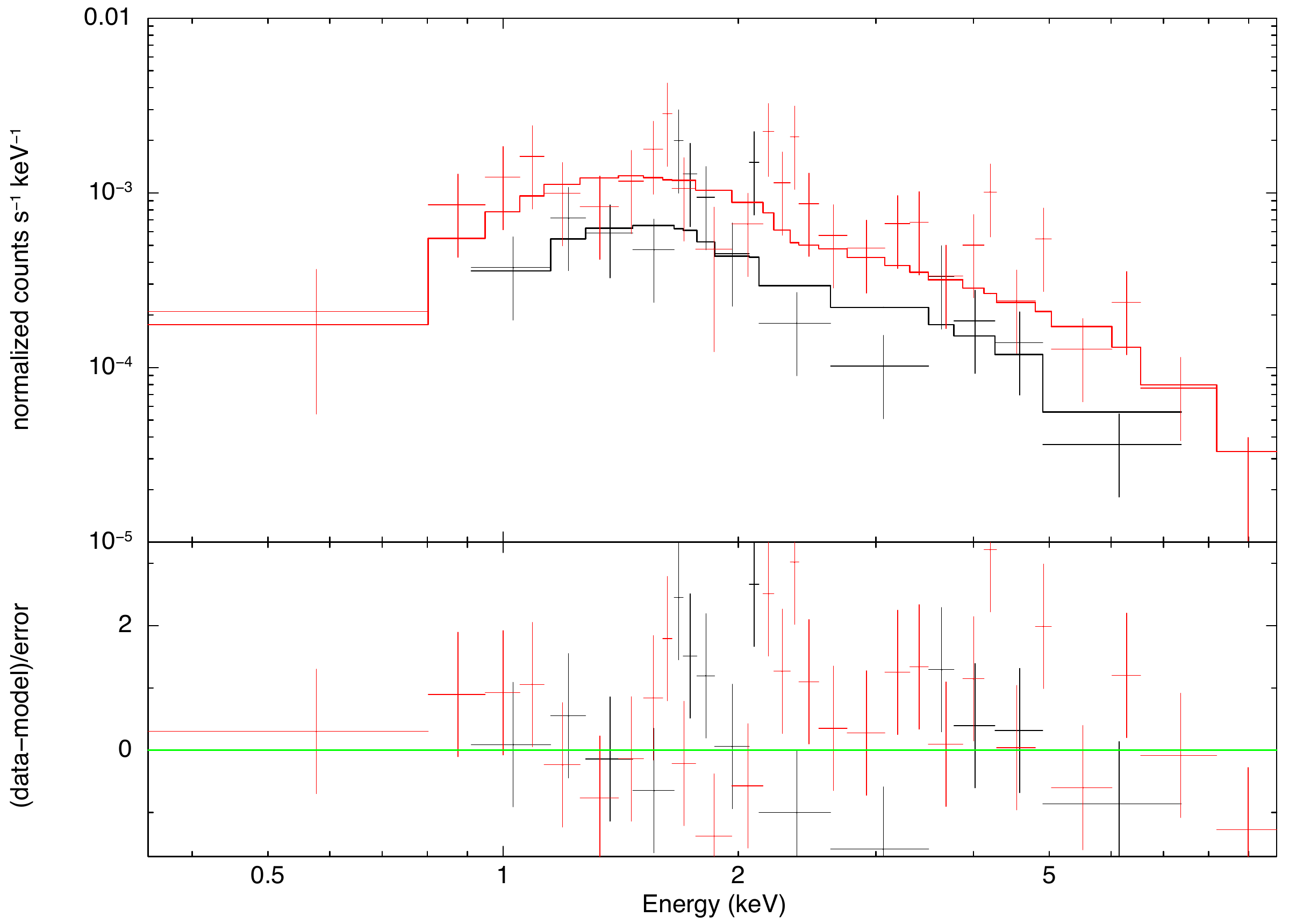}
\plotone{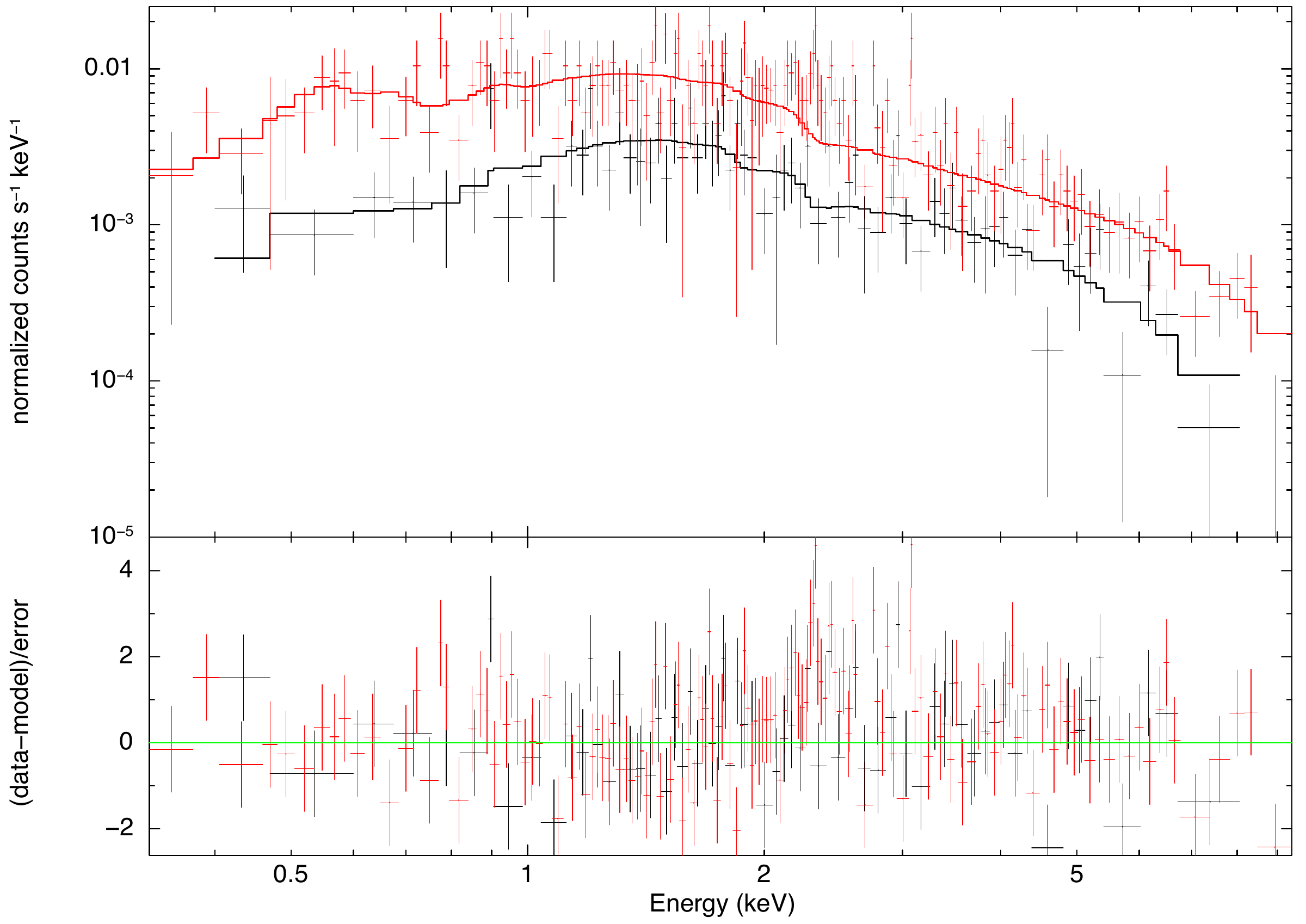}
\caption{($upper$) XMM-Newton EPIC pn and MOS2 spectra of src1 fitted with PL in 2$\arcsec$ Chandra position, and ($lower$) XMM-Newton spectra of src1 (8$\arcsec$) fitted with mekal+PL models (see text for details). \label{fig:fig3}}
\end{figure}
\begin{figure}
\plotone{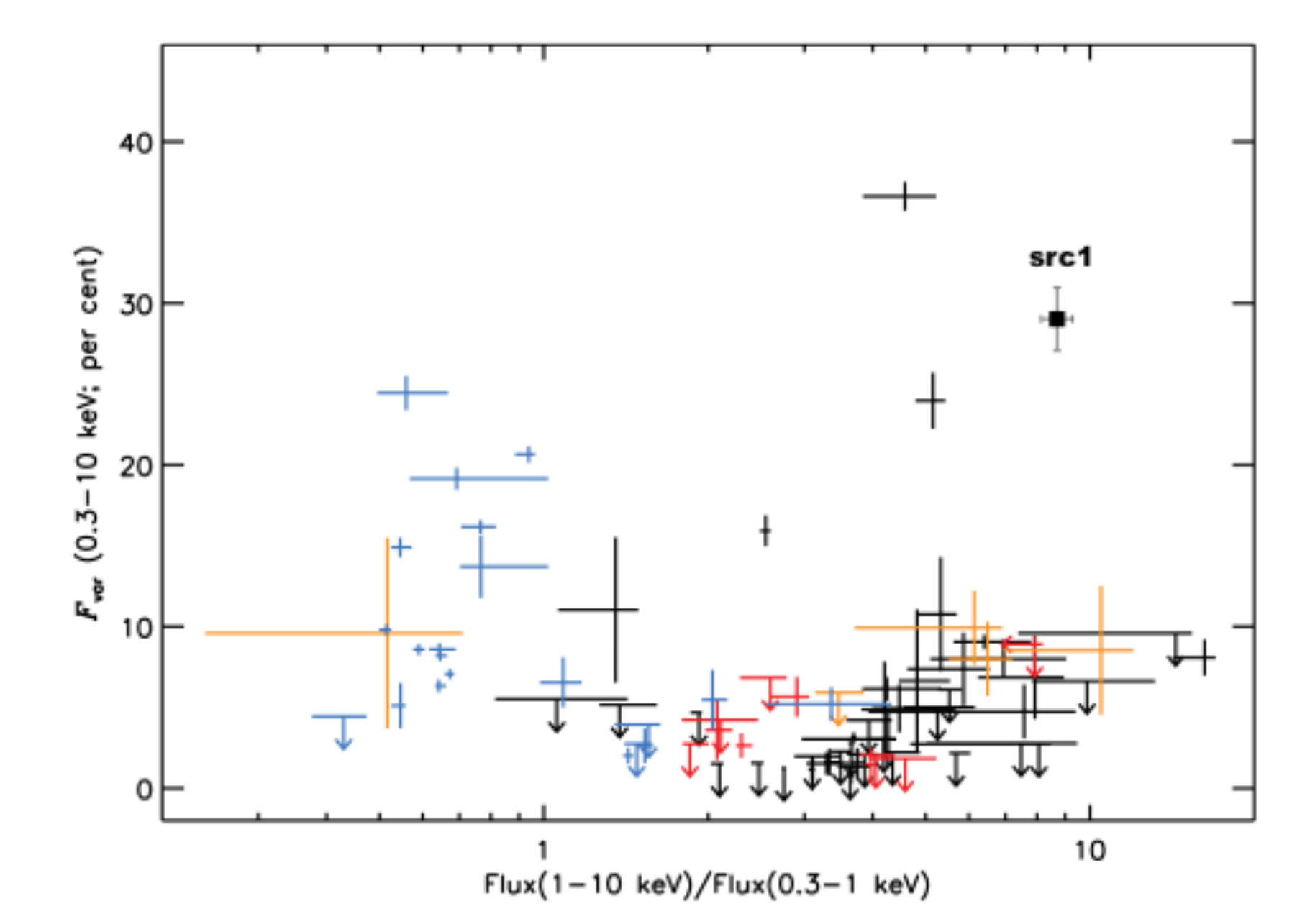}
\caption{Full-band (0.3$-$10 keV) fractional variability of a ULX sample plotted against spectral hardness. Figure reproduced from \citet{2013MNRAS.435.1758S}. Our result for src1 is indicated as a black square. \label{fig:fig4}}
\end{figure}
\\
In addition to the spectral analysis, we also performed a temporal study deploying the fractional variability as described in \citet{2003MNRAS.345.1271V}. Variations in the flux as a function of time can hint at a variability nonetheless it can also be the result of simple stochastic fluctuations that have little do with an intrinsic variability in the actual production mechanism being investigated. Fractional variability, on the other hand (with appropriate estimates of the associated errors), provides a better handle on the underlying temporal variations. If present, the scale of these temporal variations is likely to depend on the spectral state and possibly also on the spectral components that contribute to the production mechanisms and therefore one can hope, at least in an ideal situation, to use this as a diagnostic tool to not only identify the spectral state but also separate the major components. We calculated the full-band (0.3 - 10 keV) fractional variability in addition to that for the soft (0.3 - 1 keV) and hard band (1 - 10 keV) respectively. Light curves were extracted in each of the three energy bands with 200 s temporal binning, allowing us to probe variability on time-scales from the corresponding Nyquist frequency up to the full available good time in each observation. Our result is displayed (as a black square) in Figure 4 (reproduced from \citet{2013MNRAS.435.1758S}), which shows the calculated fractional variability in the full band as a function of the spectral hardness. While the uncertainty in our extraction is relatively large, the result does however suggest that the variability of src1 is significant ($\sim$30$\%$). This level of fractional variability exhibited by src1 would be rather difficult to understand in terms of an extended source with dominant diffuse emission and is much more likely associated with a point source (as exhibited for example by stellar BH binaries in the Low/Hard state). This is another reason, in addition to those noted earlier, to favor the PL+diskbb fit for src1 instead of the one that includes the diffuse emission via the Mekal model. Furthermore, we note that the variability for src1 lies in the cluster of points (black crosses, Figure 4) associated with the spectral state identified as the broadened disk, which is consistent with the results of our spectral analysis of src1. Finally, we mention that, although extremely unlikely, it is conceivable that the observed variability in the location of src1 results from a background source such as an AGN or a blazar. This possibility is readily discounted by performing a search for objects such as AGN and blazars from the available catalogs. We found no such background source within a search radius of 30$\arcsec$ from the location of src1. This null result is further confirmed by computing the number of background sources from the high-precision X-ray logN-logS distribution of AGN compiled by \citet{2008A&A...492...51M}.\\
\\
\noindent Identification of the optical counterparts of ULXs has become more of a necessity, especially in the light of recent findings that suggest that the underlying population of ULXs is likely to be heterogeneous. Pragmatically of course, both the accretion disk, via X-ray photoionization, and the donor star potentially contribute to the overall optical emission and thus each of these contributions need to be disentangled in order to place meaningful constraints on both the central compact object and the companion. Optical counterparts of a number of ULXs in nearby galaxies have been found with the use of archival Hubble Space Telescope (HST) data (\citep{2009MNRAS.397.1836G, 2011ApJ...737...81T} and references therein), but unfortunately such data are not available for NGC 3413. Some SDSS data are however available, and we make use of these data to probe the environment of the bright X-ray source near the optical center of NGC 3413. In Figure 5, we present a three-color SDSS image of NGC 3413, on which are overlaid, the positions of 6 known optical sources (white circles) that overlap the $8^{\prime\prime}$ (blue) error circles defining the positions of the X-rays sources src1 and src2, respectively. In addition, 7 other potential optical counterparts, appearing in the PanStars catalogue, are also shown ($1^{\prime\prime}$ magenta circles) to overlap the X-ray-source circles, in particular, 3 sources (magenta circles) within the green circle defining the very bright SDSS source in the optical center. The available magnitudes and the RA, Dec information for these sources is summarized in Table 3. Technically, since src1 and src2 are not resolved within the limits of the XMM resolution, any of the noted sources could represent the optical counterpart. Given the limited resolution, we make no further distinction among these optical sources. Using the available SDSS magnitudes for the optical sources, we construct a color-magnitude diagram (CMD) to estimate the age of the sources. For the CMD (shown in Figure 6), in addition to the aforementioned optical sources, stars within the 90$\%$ brightness contour, and with significant S/N ratios, are also shown as gray circles. The PARSEC isochrones14 of \citet{2012MNRAS.427..127B} were used in the CMD, based on the updated version of the code used to compute stellar tracks. The metallicity of NGC 3413 (Z = 0.0187) was determined from the mass-metallicity relation given by \citet{2004ApJ...613..898T} to obtain the isochrones. The Galactic reddening E (B - V) of this region is given from dust maps as 0.0192 according to \citet{2011ApJ...737..103S}. Based on the CMD, we estimate the age of the sources to be $<$12 Myr. We recognize this is, at best, a crude estimate given the large error circles used to select the potential counterparts, nonetheless the exercise does suggest a relatively young counterpart. Assuming the counterpart is a main-sequence object, we deploy the mass-luminosity correlation with a power-law index in the range 3-4, to obtain a mass for the potential counterpart (for src1) in the range 8 - 18M$_{\odot}$. This is consistent with the X-ray source (src1) being a high-mass X-ray binary. As eluded to above the optical center ($\sim$$3^{\prime\prime}$ from the src1) is very bright compared to the next nearest candidate i.e, a g-magnitude of 13 vs. 25. This suggests multiple unresolved sources in the optical center, a region spanning $\sim$120 pc, given the resolution of the SDSS. Indeed, the superior resolution of the PanStars indicate at least 3 sources near the optical center. As before, by deploying the SDSS magnitudes and the mass-luminosity correlation, we estimate the mass range for the stellar (and/or diffuse) sources in the optical center to be $\sim$(130 - 670)M$_{\odot}$. The sub-arcsecond resolution of the Hubble Space Telescope would be very helpful in resolving the multiplicity of the sources in the optical center of NGC3413.\\
\begin{figure}
\plotone{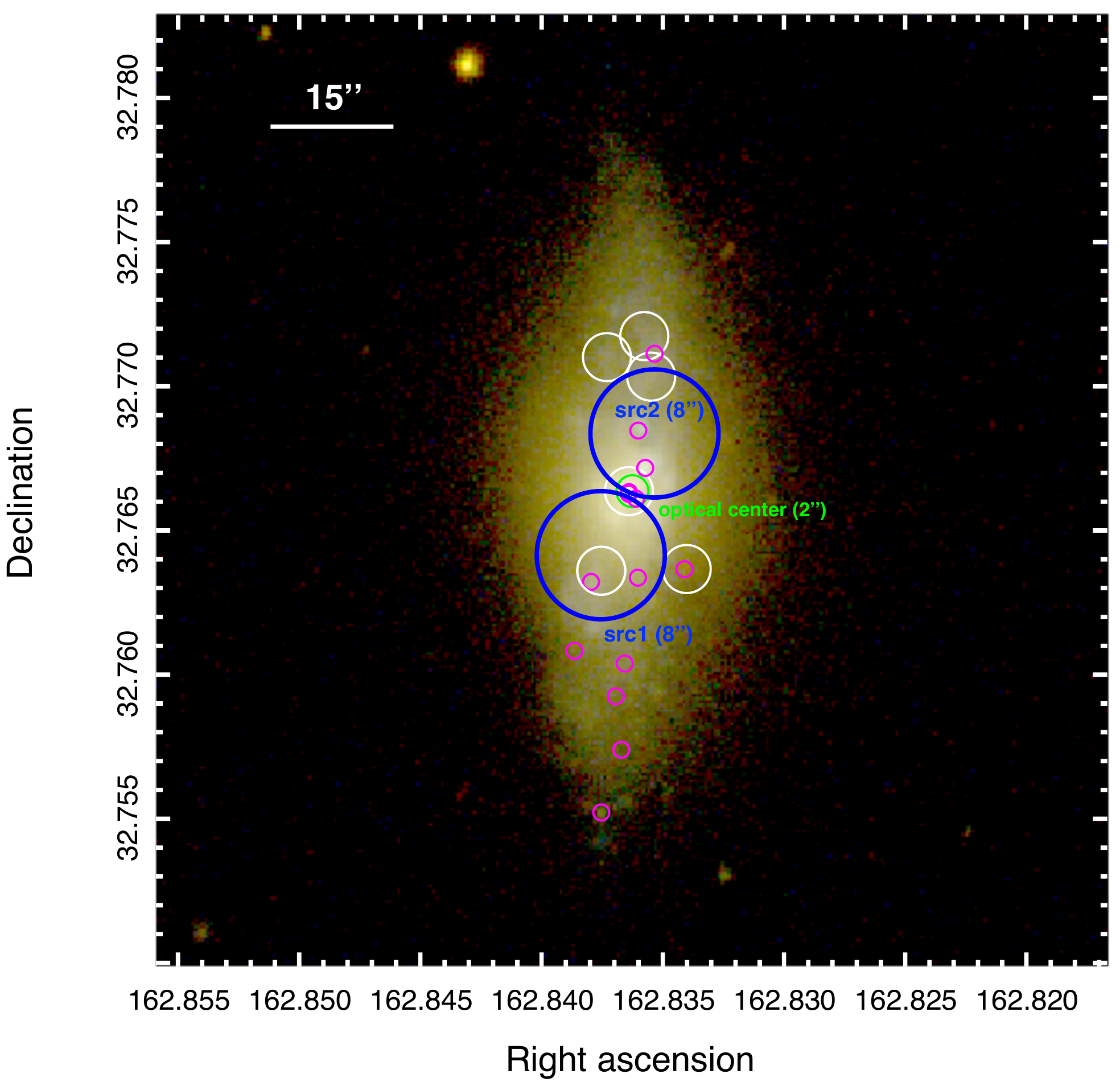}
\caption{Three-color SDSS image of NGC 3413 overlaid with blue circles (8$^{\prime\prime}$) indicating the X-ray sources src1 and src2 respectively. Potential optical counterparts from SDSS (white circles) and the PanStars catalog (magenta circles overlapping the blue circles) are also shown. The green circle indicates the optical center \label{fig:fig1}}
\end{figure}
\begin{figure}
\plotone{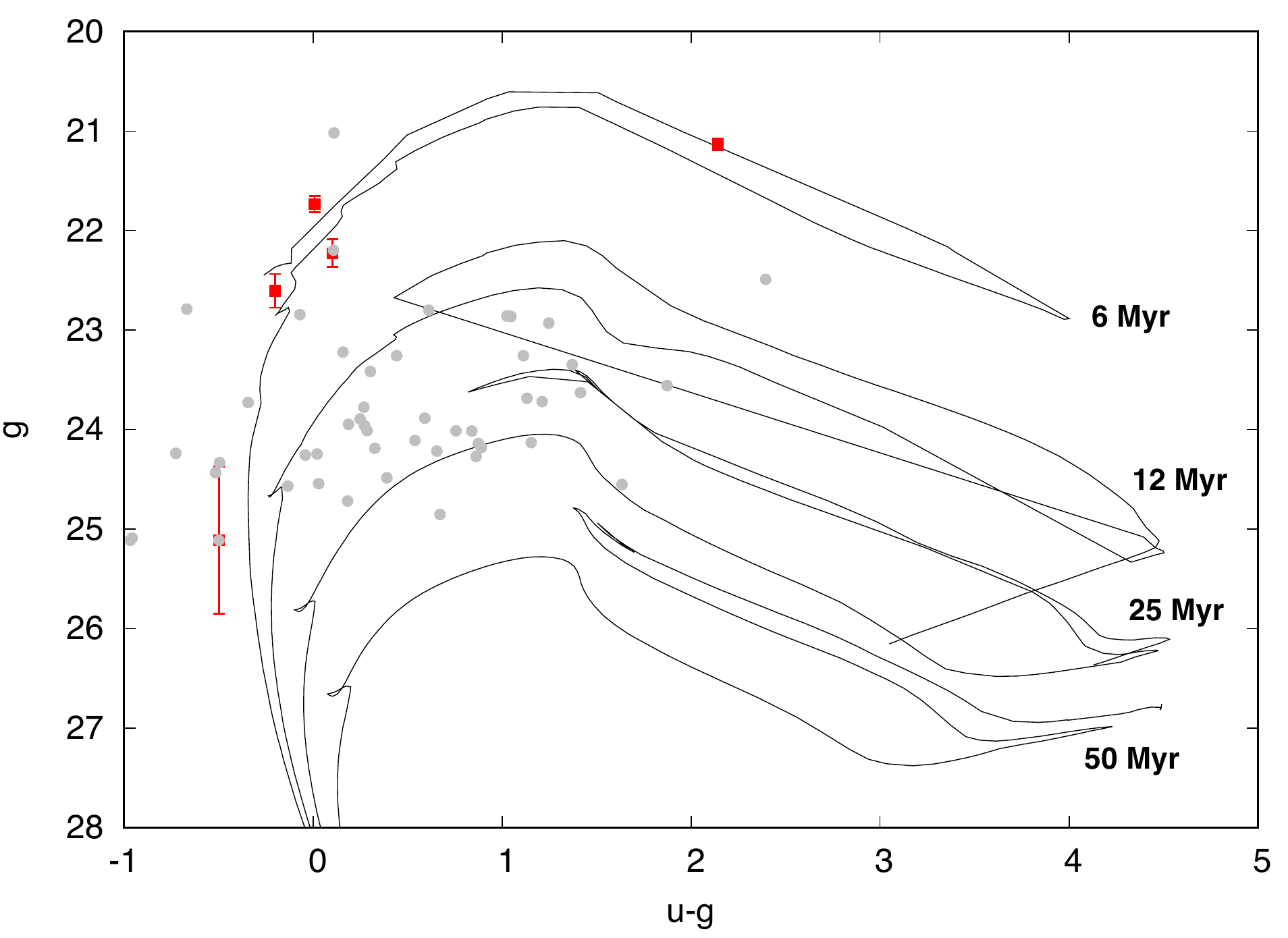}
\caption{CMD (SDSS magnitudes g vs u-g) for a number of optical sources (grey circles) within the 90$\%$ brightness contour of NGC 3413. Potential counterparts to the X-ray sources src1 and src2 are indicated in red. PARSEC isochrones in the age range 6$-$ 50 Myr are shown. \label{fig:fig3}}
\end{figure}
\section{Summary and Conclusions}
\noindent
We have performed a detailed spectral analysis of deep \emph{XMM-Newton} observations of a bright X-ray source in the nearby dwarf emission-line galaxy NGC 3413. In addition, we have explored possible optical counterparts with the available SDSS and PanStars data. Our main findings are as follows:

\begin{itemize}
\item
The bright X-ray source near the optical center of NGC 3413 is not point-like, and instead, is more likely a composite of point-like sources with extended and/or diffuse emission. This is consistent with \citet{2011ApJS..192...10L} catalog where it is flagged as a source with possible extended emission
\item
Assuming the source is composed of a brighter central component (src1) and a dimmer extended region (src2), our spectral and temporal results indicate that the bright component (src1) exhibits features similar to those associated with the so-called broadened disk state of ULXs
\item
The central bright region (src1) is best fitted with a combination of PL + diskbb models. The disk temperature is relatively high ($\sim$2 keV), a feature that is observed for other broadened disk states in ULXs. Furthermore, the fractional variability is also relatively high and is consistent with that observed for other ULXs in the broadened disk state. The turnover typically associated with ULX spectra is not immediately obvious from a visual inspection but a hint of its presence is obtained if one fits the spectrum with a broken power law; the break, while not tightly constrained, is implied around $\sim$ 3 keV. Hint of some diffuse emission is indicated with a fit using the mekal model but the extracted gas temperature is unusually low
\item
Based on the CMD for NGC 3413 and PARSEC isochrones, we estimate the age of possible optical counterparts to be $<$ 12 Myr, although we note that this is a rather crude estimate given the large error circles associated with the X-ray sources. Using the SDSS magnitudes and with the assumption of a main-sequence counterpart, we estimate its mass to be in the range 8 - 18M$_{\odot}$, making src1 consistent with a high-mass X-ray binary
\item
We follow the procedure of \citet{2000ApJ...535..632M} and use the normalization of the multi-color blackbody spectral fit to a determine a mass range of 3 - 20 M$_{\odot}$ for the compact object associated with src1
\item
The dimmer region (src2) is best fitted with a combination of mekal + Raymond models, suggesting the presence of an extended source with diffuse emission.
\item
The bright optical center, spanning a region $\sim$120pc (assuming SDSS resolution), is consistent with multiple unresolved sources with a total mass in the range $\sim$(130 - 670)M$_{\odot}$.
\end{itemize}

\noindent Finally, we note that the sub-arcsecond resolution of the HST would be very helpful in resolving the multiplicty of the optical sources in the center of NGC 3413. Similarly, a deep Chandra observation, with its superior spatial resolution, would definitely settle the issue whether the bright X-ray source is point like or a composite of several sources.

\begin{deluxetable*}{crrrrrrrr}
\tablecaption{Possible counterparts for X-ray sources src1 and src2: Optical data from $sdss$ and $panstarrs$ source catalogs.}
\tablehead{ \colhead{ObjID} & \colhead{RA} & \colhead{DEC} &\colhead{u-mag} & \colhead{g-mag}   \\
\colhead{} & \colhead{degree} & \colhead{degree} & \colhead{AB Mag.} & \colhead{AB Mag.}   }
\startdata
{\bf sdss sources} \\
J105121.00+324549.0\tablenotemark{1} & 162.838 & 32.764 &  23.29$\pm$0.68  &  21.13$\pm$0.06   \\
J105120.53+324613.2\tablenotemark{2} &  162.836 & 32.770 &  22.35$\pm$0.29  &  22.23$\pm$0.14 \\
J105120.73+324558.9\tablenotemark{3} &162.836 & 32.766 & 14.24$\pm$0.003 & 13.17$\pm$0.002   \\
{\bf panstarrs sources} \\
147311628360796000\tablenotemark{1}  &  162.836 & 32.763 & \nodata  &  \nodata  \\
147311628382496000\tablenotemark{1} & 162.838 & 32.763 & \nodata  &  \nodata    \\
147321628360082000\tablenotemark{2}  &  162.836 & 32.768 & 19.83 &  \nodata  \\
147321628359231000\tablenotemark{2} & 162.836 & 32.767 &  18.21$\pm$0.08  &  \nodata   \\
147321628354416000\tablenotemark{2} &  162.835 & 32.771 & \nodata  &  \nodata    \\
147311628362049000\tablenotemark{3} & 162.836 & 32.766 & 17.10$\pm$0.01 &  \nodata    \\
147321628364750000\tablenotemark{3} & 162.836 & 32.766 & 16.71$\pm$0.12 &  \nodata    \\
147311628365499000\tablenotemark{3} & 162.836 & 32.766 & \nodata  &  \nodata    \\
\enddata 
\label{table:nonlin} 
\tablenotetext{1}{sdss/panstarrs potential counterpart for src 1.}
\tablenotetext{2}{sdss/panstarrs potential counterpart for src 2.}
\tablenotetext{3}{sdss/panstarrs optical source near center.}
\end{deluxetable*}

\acknowledgements
.... 
\software{SAS (v14.0.0; Gabriel et al. 2004), XSPEC (v12.9.1; Arnaud 1996)}

\end{document}